\documentclass[iop]{emulateapj-rtx4} 
\shortauthors{Sekanina}
\shorttitle{Comments on Recent Review Paper}
\slugcomment{Version \today }

\begin{document}
%
\title{COMMENTS ON A RECENT REVIEW PAPER ON NEAR-SUN COMETS}
\author{Zdenek Sekanina}
\affil{Jet Propulsion Laboratory, California Institute of Technology,
  4800 Oak Grove Drive, Pasadena, CA 91109, U.S.A.}
\email{Zdenek.Sekanina@jpl.nasa.gov{\vspace{0.15cm}}}

\begin{abstract}
It is noted that some results of the author's research papers cited in the
comprehensive review paper {\it The Science of Sungrazers, Sunskirters, and
Other Near-Sun Comets\/} by G.\ H.\ Jones et al.\ (2018) have been factually
misrepresented.  The purpose of this note is to offer (i)~revisions in order
to correct the numerous errors and inaccuracies and (ii)~additions where the
text is confusing.
\end{abstract}


\section{Introduction}
A comprehensive review paper {\it The Science of Sungrazers, Sunskirters, and
Other Near-Sun Comets\/} by Jones et al.\ (2018), published recently in the {\it
Space Science Reviews\/} and referred to hereafter as {\it Paper~1\/}, was written
to summarize the recent progress in the investigation of the various categories of
comets closely approaching the Sun, including the Kreutz system of sungrazers.
As illustrated {\vspace{-0.03cm}}by the number of citations (12 by late August~of
2019, over 1$\frac{1}{2}$ years after publication,{\vspace{-0.02cm}} according to
the NASA ADS), this is an influential work, expected to offer an accurate review
of relevant research papers.

When I read {\it Paper 1\/} after its publication,\footnote{I had no opportunity
to read the paper or its draft earlier.} I noticed, to much dismay, that some
of the {\small \bf results of my cited research were factually misrepresented}.
I called the authors' attention to the shortcomings --- which should not have
passed peer review --- and received an apologetic but unsatisfactory reply, with
no commitment to fixing the errors.  Under these circumstances, issuing the
corrections below has been the only option available to~me to remedy the perceived
problems.

I focus exclusively on the citations in {\it Paper~1\/} of research results
presented in papers written by myself and my collaborators and on closely
related issues.\footnote{I employ Jones et al.'s (2018) reference identifications,
except that Sekanina (2000) now becomes Sekanina (2000a); several references in
this note are not among the references in {\it Paper~1\/}.} I have not
examined, and do not comment on, the treatment in {\it Paper~1\/} of research
results by others. Nor do I argue in this note the merits of competing hypotheses,
leaving the decision to the reader.  The purpose here is to provide revisions
and/or additions that {\it Paper~1\/} needs in order~to {\small \bf correctly
and accurately describe this research}.

%
%
\section{Post-Perihelion Appearance of\\Comet C/2011 W3}
With reference to the sungrazer C/2011 W3 (Lovejoy), the authors write incorrectly
on page~12 that {\small \bf ``[a]fter peri\-helion, the comet was only observed
as a headless~tail.''}  They essentially repeat this fallacy on page~21 using
different words, namely that {\small \bf ``no ground-based post-perihelion
observations detected a central condensation or any other indication of ongoing
activity (cf.\ Sekanina and Chodas 2012, \ldots).''}

A figure on page 3 of this cited paper, reproduced below in Figure~1, exhibits four
images of the comet taken at the Pierre Auger Observatory, Malarg\"{u}e, Argentina.
Contrary to the above account given in {\it Paper~1\/}, the {\small \bf first three
images}, taken on 2011 December~17.37--19.37~UT, that is, 1.36--3.36 days {\small
\bf after perihelion}, clearly {\small \bf do show the presence of a strong
condensation}, whose  dimensions were growing with time.  The condensation
disappeared only by the time of the fourth image, December~20.33~UT (or 4.32~days
after perihelion).  A {\small \bf spine tail}, detected already in the third
image and dominating the fourth, became the comet's brightest part from late
December on and was shown in the above cited paper to be a synchronic formation
{\small \bf composed of} millimeter-sized and larger {\small \bf grains of dust
expelled in an episode that peaked} rather sharply {\small \bf 1.6\,{\boldmath
$\pm$}\,0.2~days post-perihelion}.  No dust released preperihelion was contained
in the headless tail.

We concluded that the {\small \bf comet did survive perihelion and did
disintegrate {\boldmath $\sim$}2~days after perihelion}, not that it\, {\small
\bf m\,a\,y} \,have survived perihelion, as is our result cited on page~12 of
{\it Paper~1}.\footnote{The authors do not seem to have made up their mind on
this subject.  On page 40 they write correctly that ``\ldots \,Sekanina
and Chodas (2012) propose \ldots \,the~\mbox{disintegration} of Comet C/2011~W3
Lovejoy a day or so after perihelion.''~Yet, on page 42 one finds once again
that the comet ``may~have~survived~.\,.\,. perihelion (Sekanina and Chodas
2012, \ldots).''}.  The comet most certainly~did~not disintegrate before or at
perihelion.

\begin{figure*}[ht]
\vspace{-3.47cm}
\hspace{-0.7cm}
\centerline{
\scalebox{0.8}{
\includegraphics{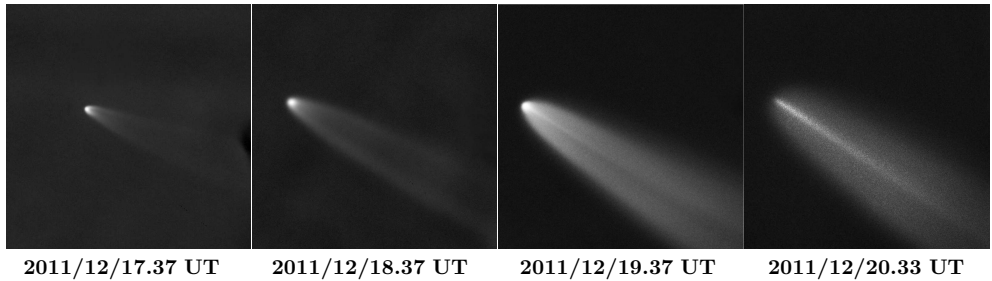}}}
\vspace{-16.19cm}
\caption{Some of the earliest postperihelion images of comet C/2011 W3 taken,
at the request of J.\ \v{C}ern\'y, by J.\ Ebr, M.\ Prouza, P.\ Kub\'anek,
and M.\ Jel\'{\i}nek with the FRAM 30-cm f/10 Schmidt-Cassegrain reflector,
a robotic, remotely controled telescope at the Pierre Auger Observatory,
Malarg\"{u}e, Argentina.  Each frame is $\sim$11$^\prime$ on a side,
equivalent to 430,000 km in the first image and 375,000~km~in~the last
iamge.  North is up, east to the left.  The comet was 5$^\circ\!$.8 from
the Sun during the first exposure, 8$^\circ\!$.5 during the second,
11$^\circ\!$.0 during the third, and 13$^\circ\!$.3 during the last one.
These images provided the first evidence of the major physical changes in
the comet's morphological appearance, which culminated with the suden,
{\vspace{-0.04cm}}complete loss of the nuclear condensation on December 20.
The comet was at perihelion on December 16.01~TT.  (Adapted from Figure~1
of Sekanina \& Chodas 2012; image credit:\ J.\ \v{C}ern\'y, Czech
Astronomical Society.)}
\vspace{0.69cm} 
\end{figure*}

\section{Fragmentation and Age of\\the Kreutz System}
%
%
The orbital evolution of the Kreutz sungrazer system, the most extensive known
group of genetically related comets, is described in a paragraph on page~23 after
it~is affirmed that {\small \bf ``most investigators \,\ldots \,are~in~\mbox{agreement} about the general picture of the group's evolution~[in which]
the \ldots \,original parent comet was perturbed into a sungrazing orbit and broke
up near perihelion.''}  As cited,\footnote{This general description is accompanied
by~one~line in Table~4 and by Figure~12.  In addition, isolated, terse, and nearly
identical references to the subject are scattered all over the paper; one learns,
for example, starting at the foot of page~3 and on the first line of page 4 that
the ``catalogued members of the population \ldots are the result of repeated
fragmentations of parent objects \ldots''  A similar sentence is on page~11,
while on page~24 it is stated that the ``Kreutz system is still evolutionarily
young,'' whatever that means.} the ``most investigators'' consist~of~Marsden
(1967, 1989) and Sekanina \& Chodas 2002a, 2002b, 2004, 2007, 2008).
Two points to emphasize are that (i)~neither Marsden nor Sekanina with Chodas
paid any attention to the process of the parent's perturbation into the
sungrazing orbit and that (ii)~ignored, unfortunately, is the eminently relevant
paper (Sekanina 2002b),\footnote{Cited in the preceding paragraph of {\it
Paper\,1} in another context.} in which it was for the first time clearly
proposed that the {\small \bf \ldots \,early breakup of the progenitor comet
itself\, \ldots \,could not have taken place near
perihelion},\footnote{Fragmentation far
from perihelion is acknowledged in {\it Paper\,1\/} on pages 23/24 and 43 only in
relation to the faint, SOHO sungrazers (Sekanina 2000a, 2002a).} in contradiction
to what is stated in {\it Paper~1}.  The notion of the initial breakup far from
the Sun was then developed in the framework of cascading fragmentation into two
different scenarios by Sekanina \& Chodas (2004, 2007), with either C/1882~R1 or
C/1843~D1 assumed to contain the major residual mass of comet X/1106~C1.  The three
remaining \mbox{2002--2008} papers cited in {\it Paper~1\/} pertain to specific
sungrazers and are far less relevant in this context. 

%
The idea of the initial fragmentation episode at {\it large\/}
heliocentric distance has major ramifications because it (i)~allows
for the sizable differences among the Kreutz sungrazers in both
the angular elements and perihelion distance (cf.\ Section~4) as
products of an instantaneous event (or events); and (ii)~does not
contradict the notion, consistent with independent evidence, that
the Kreutz system originated very recently, not more than two
or (at most) three revolutions about the Sun ago.  On the other
hand, the hypothesis of the initial fragmentation episode at {\it
perihelion\/},{\vspace{-0.04cm}} independently advocated by both
\mbox{Marsden} (1967) and \"{O}pik (1966), has to resort to the indirect
planetary perturbations as the trigger for the differences in the
orbital elements.  And because their effect per revolution is
substantially smaller than the observed disparity (Marsden 1967,
1989), the number of required orbits about the Sun is accordingly
higher.

Harwit (1966) concluded that the breakup process of the type
undergone by comet Ikeya-Seki (C/1965~S1), i.e., at perihelion,
cannot account for some of the orbital differences, although
Marsden (1967) disagreed.  On the other hand, Marsden did briefly
consider fragmentation far from the Sun as a solution to the
problem, but dismissed it unless an explanation was provided for
such an event ``when the velocity of separation is some 20\% of
the velocity of the comet itself!''  It is true that the mechanism
of nontidal fragmentation far from the Sun is a controversial
subject, yet the evidence that this process takes place is overwhelming.

I should add that the developed fragmentation model possesses a limited
prognostication quality.  Analysis of the bright Kreutz sungrazers' clusters
over the past five
centuries as a product of cascading fragmentation allowed a prediction by
Sekanina \& Chodas (2007) of the anticipated arrival of {\small \bf ``another
cluster \ldots \,in the \mbox{coming} decades, with the earliest
objects expected \ldots \,as soon as several years from now.''}  The
discovery of comet Lovejoy (C/2011~W3) four years following this prediction
was its {\small \bf spectacular confirmation,} as seen from an updated
plot in Figure~2. (No reference to this important, validated prognosis is
made in {\it Paper~1\/}.){\hspace{0.3cm}}

These are elements of the kind of progress summary of the Kreutz group
evolution research that {\it Paper~1\/}~should have provided.  The
reader looking for such an account should turn to another source.
One is Marsden's (2005) in-depth review paper, which though becomes
somewhat outdated.  To a degree, it has been complemented by the
relevant chapters of Seargent's (2012) meticulously researched book,
which, while aiming at a wider readership, is a good compendium of
recent results at all levels.

\begin{figure*}
\vspace{-4.45cm}
\hspace{0cm}
\centerline{
\scalebox{0.85}{
\includegraphics{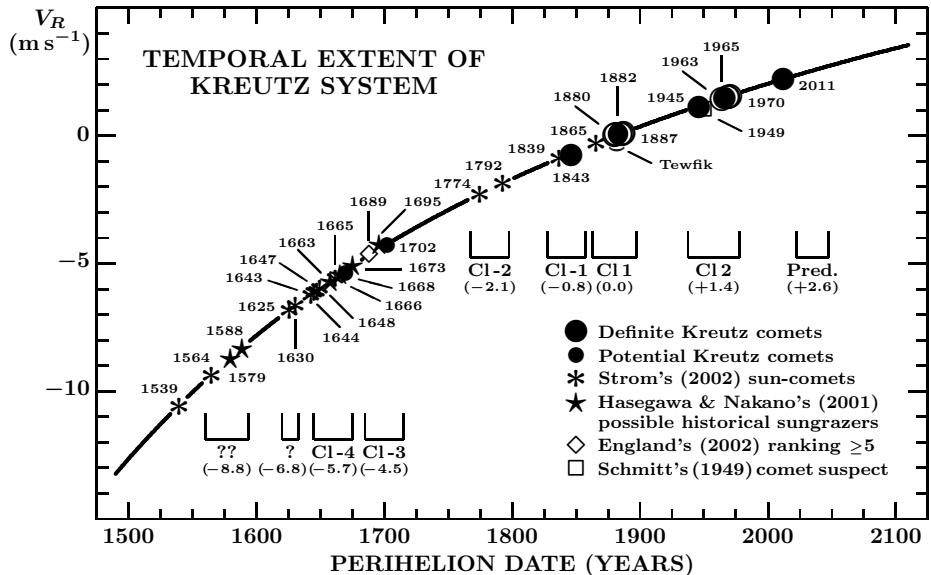}}}
\vspace{-13.45cm}
\caption{Temporal extent and apparent clustering of the Kreutz system.  The
{\vspace{-0.01cm}}various symbols used for the sungrazers are explained~in the
figure.  The radial component of the separation velocity, $V_R$ (in m~s$^{-1}$),
needed in the course of an assumed fragmentation event of comet X/1106~C1 that
{\vspace{-0.055cm}}occurs about 2~hours after perihelion to explain the time
gap from C/1882~R1 (for which \mbox{$V_R = 0$}), is plotted as a function of
perihelion date.  For example, less than +2~m~s$^{-1}$ is needed for a perihelion
arrival in 1965 (as did comet Ikeya-Seki) instead in 1882.  The time intervals
apparently occupied by the various clusters (Cl\,2, Cl\,1, etc.) are depicted by
the horizontal brackets, with the representative values of $V_R$ in parentheses.
{\sf Pred.}\,is the time span predicted for the core of the 21st century cluster.
A similar plot can be constructed for the transverse component of the separation
velocity.  Adapted from Figure~4 of Sekanina \& Chodas (2007) and updated to
include comet C/2011~W3 (Lovejoy), the first object signalling the forthcoming
arrival of the predicted new cluster in the 21st century.{\vspace{0.8cm}}}
\end{figure*}

\section{Orbital Similarity}
While it is possible to say that Kreutz~\mbox{sungrazers}~move about the Sun in
generally similar orbits,~one~should~{\small \bf not equate} the attribute {\small
\bf similar} with {\small \bf nearly identical} to the point of ignoring the orbital
differences~among~the~individual members of the Kreutz system altogether.~Yet,
precisely {\small \bf this is done} in {\it Paper~1\/}.  The notion of two
subgroups, I and II, strongly promoted by \mbox{Marsden} (1967, 2005), is never
mentioned in {\it Paper~1\/}, even though it en\-tails differences of up to
21$^\circ$\,in the angular
elements.~In\-deed, Marsden (1967) remarked, on page 1171, that these elements
and the perihelion distance of eight bright members of the system {\small \bf
``differ quite markedly."}~Similarly, Sekanina \& Chodas (2004) referred, on
page 624,~to~the {\small \bf ``sizable differences between the angular
elements."} The discovery of C/1970~K1 widened this range for~the sungrazers
with well-determined orbits from 21$^\circ$ to 31$^\circ$, compelling Marsden
(1989) to introduce subgroup IIa, while the arrival of C/2011~W3 one year after
Marsden passed away extended the range among~the~Kreutz~com\-ets observed from the
ground by yet another 11$^\circ$ to 42$^\circ$!  A wide {\small \bf interval} of
the longitudes of the ascending node of the 1843--2011 bright sungrazers {\small
\bf from 326{\boldmath $^\circ$} to 8{\boldmath $^\circ$} does not compare
favorably with the ``typical"~value}~of~0$^\circ$ in Table~4 of {\it Paper\,\,1}
or, for that matter,~with any other constant value.
The~SOHO~\mbox{members}~of~the Kreutz~sys\-tem, whose
scatter in the argument of perihelion in Figure~12 (failing,
incidentally, to include C/2011~W3) amounts to at least 80$^\circ$, merely add
insult to injury.  The only orbital property that all Kreutz sungrazers
share with high accuracy is the {\small \bf common direction of the line
of apsides}, which was already noted, in a somewhat cryptic way, by Kreutz (1895)
to apply to comets C/1843~D1 and C/1882~R1.

The arbitrary standard for orbital similarity tolerated in {\it Paper~1\/} is
illustrated by comparing the scatter among the Kreutz sungrazers in Figure~12
with the orbital differences between the Marsden and Kracht groups in Table~4.
While the {\small \bf Kreutz comets} are deemed, on page 11, to possess {\small
\bf ``very similar orbital elements,"} one learns, on page 25, that the {\small
\bf ``Marsden and Kracht orbits are currently dissimilar,"} even though the
angular elements of the two compact groups do not differ by more than
35$^\circ$ and their perihelion distances overlap.

The ambiguous definition of what orbits are or are not similar is bound to
confuse the reader, and so is the failure to pursue an issue to the end.
Staying with Figure~12, one immediately notes in the left panel that the
manner in which the orbit inclination correlates with the argument of
perihelion is very different for the Kreutz sungrazers observed from the
ground on the one hand and for the minor, SOHO members of the system on the
other hand.  The trend in the plot is from the lower left to the upper
right for the former, but mostly from the upper left to the lower right
for the latter.  Although a solution to this puzzle was proposed by Sekanina
\& Kracht (2015), no explanation is offered in {\it Paper~1}.

\section{Dust Tails of SOHO Sungrazers and\\the Lorentz Force}
On page 47/48, a strange conclusion is reached~by~the authors on the dust tails
of SOHO sungrazers;~they comment somewhat obliquely and with a wrong citation
that {\small \bf ``observations haven't yet} (sic!) {\small \bf revealed the
influence of the Lorentz force on the dust~(\mbox{Sekanina}~2000)."} I did
indeed rule out perceptible Lorentz force~effects~from diagnostic orientations
of {\it prominent\/}~narrow~tails~of~two Kreutz comets in several images taken
with the SOHO's C2 coronagraph (Sekanina 2000b, missing from the list of
references in {\it Paper~1\/}).  Even though confirmed~by Thompson's (2009)
major study (\mbox{ignored}~in~\mbox{\it Paper~1\/})~of a 2007 bright SOHO/STEREO
 sungrazer, the authors do insist that the
{\small \bf ``influence must \ldots \,be significant, as the Lorentz
force has to be invoked to explain some dust tails \ldots \,far from the
Sun (Kramer et~al.~2014)."}  ``Some dust tails" were three noisy images of a
comet~at distances between 20~AU and 30~AU from the Sun taken by the
Spitzer Telescope.  In the paper, Kramer~et~al.\ admitted that the Lorentz
force was in their opinion only the ``most likely" among several explored
explanations and, if correct, theirs was the first such detection ever reported.
Yet, this inherently weak piece of evidence is in {\it Paper~1\/} deemed more
credible and relevant to the SOHO comets than the evidence predicated~on~the~SOHO
data themselves.~The reader must be baffled by this argument.{\hspace{0.05cm}}

\section{Sunskirting Comet C/2002 R5 (SOHO)\\and Its Fragments}
On page 27 the authors write that according to Kracht
and Sekanina (Kracht et al.\ 2008), C/2002~R5 split into
C/2008~L6 and C/2008~L7 but {\small \bf ``none were observed
at what would have been their next return in 2014 so the
linkage remains uncertain.''}  This is manifestly incorrect,
since {\small \bf both fragments were detected in 2014 as
SOHO-2673 and SOHO-2712} by Z.~Xu, according to 
the {\it Sungrazer Project\/} website.\footnote{See the website {\tt
https://sungrazer.nrl.navy.mil/index.php?
p=comets.found.old/comets.arch.}  The messages are dated
2014 April~17, 08:55:55 for SOHO-2673 and 2014 August 22, 15:39:47~for
SOHO-2712.} 

In his updated orbital computations Kracht was able to link C/2002~R5 with
C/2008~L6 and SOHO-2712 as well as 
with C/2008~L7 and SOHO-2673.  He also predicted that
the pair should return on 2019 November 21, the formerly trailing fragment
SOHO-2712 now ahead of SOHO-2673 by more than
6~hours.\footnote{See the website {\tt http://www.rkracht.de/soho/kracht2.htm}.{\vspace{1.5cm}}}

\section{Sporadic Sungrazers}
Also on page 27 it is suggested that {\small \bf ``for sporadic sungrazers, the
Oort Cloud is the likely origin because the inclinations of these objects are
randomly scattered across the sky.''}  Random inclinations are not a diagnostic
attribute of comets arriving from the Oort Cloud, as comets whose orbital
periods are shorter by as much as three orders of magnitude also
exhibit essentially random inclinations.  In general, the Oort Cloud origin
is possible, but not likely, for sporadic near-sun comets detected before
perihelion, but it should be rejected for such comets observed after
perihelion.  Since the objects in this non-group category are usually quite
faint intrinsically, the {\it Bortle\/} (1991) {\it limit\/} (mentioned in
another context in Section 5.2 but not here) --- and the improved test, the
{\it synoptic index for perihelion survival\/} (Sekanina 2019) --- both {\small
\bf rule out the Oort Cloud~\mbox{origin} for intrinsically faint, yet surviving
objects.}  A probable exception is comet C/2015~D1 (SOHO), observed
to disintegrate shortly after perihelion (Hui et al.\ 2015).

\section{Other Issues}
I am fully unaware that {\small \bf ``Sekanina and Kracht (2014) believe that
the nucleus [of comet ISON] underwent continuous erosion inward of $\sim$\,1~AU},''
as is stated on page~36.  This does not sound right given that the
cited paper has a special subsection (5.4, starting on page 30) on
the comet's activity (and therefore erosion) far from the Sun.
In fact, Fugure~2 on page 3 of that paper shows that the
Oort Cloud comet was {\small \bf already active at 9.4~AU from the Sun},
during the first of the five production cycles, when the
intrinsic brightness varied as an inverse fifth power of
heliocentric distance.

The reference, on page 48, to Sekanina (2000) (i.e., to Sekanina 2000a below,
see footnote~2) in relation to a study of striated tails of comets is incorrect;
the correct reference, Sekanina \& Farrell (1980), is on page 49.

In Sekanina \& Chodas (2012), we did not arrive at any conclusion that could be
understood as indicating that {\small \bf ``C/2013~W3 Lovejoy survived well within
the Roche limit with only partial fragmentation (Sekanina and Chodas 2012),''} as
is maintained on page 62.  On the contrary, we addressed, on page 21 of the cited
paper, a question of {\small \bf ``[w]hat property was it \ldots \,that made
C/2011~W3 survive perihelion on the one hand but suddenly disintegrate nearly
two days later on the other hand.''}  We saw no evidence for or against the
nucleus having been ``partially'' fragmented at perihelion and did not get
involved with that issue at all.\\[0.1cm]

This work was carried out at the Jet Propulsion Laboratory, California
Institute of Technology, under contract with the National Aeronautics and
Space~Administration.\\[0.1cm]
\begin{center}
{\footnotesize REFERENCES}
\end{center}
\vspace{-0.3cm}
\begin{description}
{\footnotesize
\item[\hspace{-0.3cm}]
Bortle, J.\ E.\ 1991, Int.\ Comet Quart., 13, 89
\\[-0.57cm]
\item[\hspace{-0.3cm}]
England, K.\ J.\ 2002, JBAA, 112, 13
\\[-0.57cm]
\item[\hspace{-0.3cm}]
Harwit, M.\ 1966, AJ, 71, 857
\\[-0.57cm]
\item[\hspace{-0.3cm}]
Hasegawa., I., \& Nakano, S.\ 2001, PASJ, 53, 931
\\[-0.57cm]
\item[\hspace{-0.3cm}]
Hui, M.-T., Ye, Q.-Z., Knight, M., et al.\ 2015, ApJ, 813, 73
\\[-0.57cm]
\item[\hspace{-0.3cm}]
Jones, G.\ H., Knight, M.\ M., Battams, K., et al.\ 2018, Space Sci.{\linebreak}
 {\hspace*{-0.6cm}}Rev., 214, 20
\\[-0.57cm]
\item[\hspace{-0.3cm}]
Kracht, R., Marsden, B.\ G., Battams, K., \& Sekanina, Z.\ 2008,{\linebreak}
 {\hspace*{-0.6cm}}IAUC 8983
\\[-0.57cm]
\item[\hspace{-0.3cm}]
Kramer, E.\ A., Fernandez, Y.\ R., Lisse, C.\ M., et al.\ 2014, Icarus,{\linebreak}
 {\hspace*{-0.6cm}}236, 136
\\[-0.57cm]
\item[\hspace{-0.3cm}]
Kreutz, H.\ 1895, Astron.\ Nachr., 139, 113
\\[-0.57cm]
\item[\hspace{-0.3cm}]
Marsden, B.\ G.\ 1967, AJ, 72, 1170
\\[-0.57cm]
\item[\hspace{-0.3cm}]
Marsden, B.\ G.\ 1989, AJ, 98, 2306
\\[-0.57cm]
\item[\hspace{-0.3cm}]
Marsden, B.\ G.\ 2005, Ann.\ Rev.\ Astron.\ \& Astrophys., 43, 75
\\[-0.57cm]
%
%
\item[\hspace{-0.3cm}]
\"{O}pik, E.\ J.\ 1966, Ir.\ Astron.\ J., 7, 141
\\[-0.57cm]
\item[\hspace{-0.3cm}]
Schmitt, A.\ 1949, IAUC 1221
\\[-0.57cm]
\item[\hspace{-0.3cm}]
Seargent, D.\,A.\,J.\ 2012, Sungrazing Comets:\ Snowballs in the~Fur-{\linebreak}
 {\hspace*{-0.6cm}}nace.  Amazon Books (Kindle Edition, 128pp)
\\[-0.57cm]
%
%
\item[\hspace{-0.3cm}]
Sekanina, Z.\ 2000a, ApJ, 542, L147
\\[-0.57cm]
\item[\hspace{-0.3cm}]
Sekanina, Z.\ 2000b, ApJ, 545, L69
\\[-0.57cm]
\item[\hspace{-0.3cm}]
Sekanina, Z.\ 2002a, ApJ, 576, 1085
\\[-0.57cm]
\item[\hspace{-0.3cm}]
Sekanina, Z.\ 2002b, ApJ, 566, 577
\\[-0.57cm]
\item[\hspace{-0.3cm}]
Sekanina, Z.\ 2008, IAUC 8983
\\[-0.57cm]
\item[\hspace{-0.3cm}]
Sekanina, Z.\ 2019, eprint arXiv:1903.06300
\\[-0.57cm]
\item[\hspace{-0.3cm}]
Sekanina, Z., \& Chodas, P.\ W.\ 2002a, ApJ, 581, 760
\\[-0.57cm]
\item[\hspace{-0.3cm}]
Sekanina, Z., \& Chodas, P.\ W.\ 2002b, ApJ, 581, 1389
\\[-0.57cm]
%
%
%
%
%
\item[\hspace{-0.3cm}]
Sekanina, Z., \& Chodas, P.\ W.\ 2004, ApJ, 607, 620
\\[-0.57cm]
\item[\hspace{-0.3cm}]
Sekanina, Z., \& Chodas, P.\ W.\ 2007, ApJ, 663, 657
\\[-0.57cm]
\item[\hspace{-0.3cm}]
Sekanina, Z., \& Chodas, P.\ W.\ 2008, ApJ, 687, 1415
\\[-0.57cm]
\item[\hspace{-0.3cm}]
Sekanina, Z., \& Chodas, P.\ W.\ 2012, ApJ, 757, 127
\\[-0.57cm]
\item[\hspace{-0.3cm}]
Sekanina, Z., \& Farrell, J.\ A.\ 1980, AJ, 85, 1538
\\[-0.57cm]
\item[\hspace{-0.3cm}]
Sekanina, Z., \& Kracht, R.\ 2014, eprint arXiv:1404.5968
\\[-0.57cm]
\item[\hspace{-0.3cm}]
Sekanina, Z., \& Kracht, R.\ 2015, ApJ, 801, 135
\\[-0.57cm]
\item[\hspace{-0.3cm}]
Strom, R.\ 2002, A\&A, 387, L17
\\[-0.65cm]
\item[\hspace{-0.3cm}]
Thompson, W.\ T.\ 2009, Icarus, 200, 351}
\vspace{-2.3cm}
\end{description}
\end{document}